\documentclass[aps,prl,twocolumn,showpacs,superscriptaddress,floatfix,10pt]{revtex4-1}

\usepackage{graphicx}
\usepackage{amsmath}
\usepackage{epsfig}
\usepackage{helvet}
\usepackage{amssymb}

\newcommand{\ket}[1]{\mbox{$| #1 \rangle$}}

\newcommand{\proj}[1]{\mbox{$| #1 \rangle\langle #1 |$} }

\newcommand{\tr}{\mbox{$\text{tr}$}}

\def\fp{*}

\def\odd{\mbox{\scriptsize odd}}
\def\even{\mbox{\scriptsize even}}
\def\todd{\mbox{\tiny odd}}
\def\teven{\mbox{\tiny even}}
\def\fp{\mbox{\scriptsize fp}}

\begin{document}

\title{Tensor network renormalization yields \\
the multi-scale entanglement renormalization ansatz}

\author{G. Evenbly}
\affiliation{Department of Physics and Astronomy, University of California, Irvine, CA 92697-4575 USA}
\email{gevenbly@uci.edu}
\author{G. Vidal}
\affiliation{Perimeter Institute for Theoretical Physics, Waterloo, Ontario N2L 2Y5, Canada}  \email{gvidal@perimeterinstitute.ca}
\date{\today}

\begin{abstract}
We show how to build a multi-scale entanglement renormalization ansatz (MERA) representation of the ground state of a many-body Hamiltonian $H$ by applying the recently proposed \textit{tensor network renormalization} (TNR) [G. Evenbly and G. Vidal, arXiv:1412.0732] to the Euclidean time evolution operator $e^{-\beta H}$ for infinite $\beta$. This approach bypasses the costly energy minimization of previous MERA algorithms and, when applied to finite inverse temperature $\beta$, produces a MERA representation of a thermal Gibbs state. Our construction endows TNR with a renormalization group flow in the space of wave-functions and Hamiltonians (and not merely in the more abstract space of tensors) and extends the MERA formalism 
to classical statistical systems.
\end{abstract}


\maketitle

Consider a strongly interacting quantum many-body system in $D$ spatial dimensions described by a microscopic Hamiltonian $H$. Understanding its collective, low energy behavior is a main goal in condensed matter and high energy physics, one that poses a formidable theoretical challenge. To tackle this problem, plenty of methods have been proposed based on the \textit{renormalization group} (RG) \cite{Kadanoff, Wilson, Fisher}, that is, on studying how the physics depends on the scale of observation. Weakly interacting systems can be addressed perturbatively using momentum space RG \cite{Wilson}. Instead, strongly interacting systems often require non-perturbative, \textit{real space} RG methods, as pioneered by Kadanoff \cite{Kadanoff} and Wilson \cite{Wilson}.

Improving on Kadanoff and Wilson's proposals, White's \textit{density matrix renormalization group} (DMRG) for quantum spin chains \cite{DMRG} established how to systematically preserve the ground state wave-function during real-space coarse-graining -- namely, by preserving the support of its reduced density matrix. Similarly, Levin and Nave's \textit{tensor renormalization group} (TRG) \cite{TRG} taught us how to coarse-grain Euclidean path integrals of one-dimensional quantum systems (also partition functions of two-dimensional classical systems). Both DMRG and TRG are very successful, versatile approaches. However, they depart significantly from the spirit of the RG, in that they produce a coarse-grained, effective description of the system that still retains some irrelevant microscopic details. As a result, these methods (i) fail to define a proper RG flow, one with e.g. the correct structure of fixed points; and (ii) struggle to deal with critical systems or systems in $D\geq 2$ dimensions, where the accumulation of irrelevant microscopic degrees of freedom is more significant and harmful. 

These difficulties have been solved with two closely related proposals. First, \textit{entanglement renormalization} was put forward to address the above two problems in the context of ground state wave-functions \cite{ER,MERA}. By introducing \textit{disentanglers}, which remove short-range entanglement, a proper RG flow is generated, as well as an RG transformation that is computationally sustainable even at criticality. In addition, entanglement renormalization leads to an efficient tensor network description of ground states for critical systems, the \textit{multi-scale entanglement renormalization ansatz} (MERA) \cite{MERA}, of interest not only as a many-body variational state \cite{MERA, algorithms1, algorithms2, criticality1,criticality2,criticality3,criticality4, Impurity} but also as a lattice realization of the holographic principle \cite{holography, holography2}. 

\textit{Tensor network renormalization} (TNR), on the other hand, was more recently proposed to tackle the same problems in the context of Euclidean path integrals (and classical statistical systems) \cite{TNR}. The Euclidean path integral $Z \equiv \tr \left( e^{-\beta H} \right)$ is represented by a tensor network, consisting of copies of a single tensor $A$ \cite{nonTI}, which extends both in space and Euclidean time directions. Through local manipulation of this tensor network, TNR produces a sequence of tensors,  
\begin{equation} \label{eq:TensorFlow}
A \rightarrow A' \rightarrow A'' \rightarrow \cdots \rightarrow A_{\fp},
\end{equation}
corresponding to increasing length scales, which flow towards some infrared fixed-point tensor $A_{\fp}$. The later retains only the universal features of the phase or phase transition \cite{TNR}. Once again, the key of the approach is the removal of short-range correlations by disentanglers. 

In this Letter we establish a close connection between the two approaches. We show that, when applied to the Euclidean path integral \textit{restricted to the upper half plane}, TNR generates a MERA for the ground state of Hamiltonian $H$. More generally, TNR also produces a MERA for the thermal Gibbs state $\rho_{\beta} \equiv e^{-\beta H}/Z$ at finite inverse temperature $\beta$, as well as for the low energy eigenstates of $H$ on a finite periodic chain. 
Our result provides an alternative route to the MERA, one that bypasses the costly energy minimization of previous algorithms \cite{algorithms1,algorithms2} and has several other significant advantages, both conceptual and computational, that we also discuss.
Among those, we emphasize that (i) we obtain the first correct MERA representation of the thermal state $\rho_{\beta}$ \cite{HoloThermal}, together with an algorithm to find it; (ii) the connection implies that TNR produces an RG flow in the space of wave-functions and Hamiltonians, and not just in the abstract space of tensors, Eq. \ref{eq:TensorFlow}; (iii) the MERA formalism can be applied now also to classical statistical systems.
For simplicity, we consider a translation invariant system in $D=1$ dimensions, although the key results generalize to inhomogeneous systems in dimension $D \geq 1$.  

\textit{Tensor network for Euclidean time evolution.---} 
Given a translation invariant Hamiltonian $H$ in one dimension, we use a standard procedure \cite{AppendixA} to produce a two-dimensional tensor network representation of the Euclidean time operator $e^{-\beta H}$ or Euclidean path integral $\tr \left( e^{-\beta H} \right)$. This tensor network is made of copies of a single tensor $A$. If both the system size $L$ and the inverse temperature $\beta$ are infinite, then the network spans the entire ($x,\tau$)-plane, where $x$ and $\tau$ label space and Euclidean time, respectively. Here we will consider tensor networks for $e^{-\beta H}$ on three different geometries, obtained by introducing a horizontal cut at $\tau = 0$ and by choosing $L$ and $\beta$ to be either finite or infinite \cite{AppendixB}.

\begin{figure}[!t]
\begin{center}
\includegraphics[width=8cm]{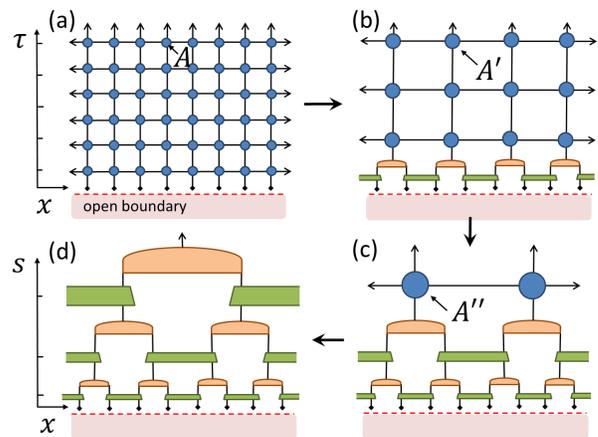}
\caption{
(a) Tensor network the ground state $\ket{\Psi}$ of $H$ on an infinite lattice. It is made of copies of tensor $A$ and restricted to the upper half plane $(x,\tau^{+})$, with a row of open indices at $\tau = 0$.
(b) By coarse-graining the tensor network while leaving the open indices untouched, we obtain a new tensor network with tensors $A'$ together with one row of disentaglers and isometries.
(c) Further coarse-graining of the tensor network produces new coarse-grained tensors $A''$ and a second layer of disentanglers and isometries.
(d) By iteration we obtain a full MERA approximation for state $\ket{\Psi}$.
} 
\label{fig:UpperHalfPlane}
\end{center}
\end{figure}

\textit{TNR yields MERA.---} Let us start with the upper half plane, Fig. \ref{fig:UpperHalfPlane}, which corresponds to the ground state $\ket{\Psi}$ of $H$ on an infinite lattice. The network has an open boundary at $\tau=0$, with an infinite row of open indices, one for each site of the one-dimensional lattice on which $H$ acts. We first apply TNR everywhere on the upper half plane except near $\tau=0$, where we keep the open indices of the tensor network untouched. TNR acts through an intrincate sequence of local replacements \cite{TNR}. Here we skip the technical details \cite{AppendixC} and focus instead on describing the final result: a coarse-grained tensor network with effective tensor $A'$ for most of the upper half plane, in accordance with Eq. \ref{eq:TensorFlow}, together with a double row of special tensors, so-called disentanglers and isometries, which correspond to one layer of the MERA \cite{layer}. These tensors connect the microscopic degrees of freedom at length scale $s=0$ (represented by the original open indices of the network) with the coarse-grained degrees of freedom at length scale $s=1$ (represented by the lower indices of the lowest row of tensors $A'$). We can now repeat the process on tensors $A'$, to obtain coarse-grained tensors $A''$ and a second row of disentanglers and isometries, i.e. a second layer of the MERA, connecting scales $s=1$ and $2$. Iteration then produces a full MERA approximation for the ground state $\ket{\Psi}$ of $H$, encompassing all length scales $s=0,1,2,\cdots$.  

\begin{figure}[!t]
\begin{center}
\includegraphics[width=8cm]{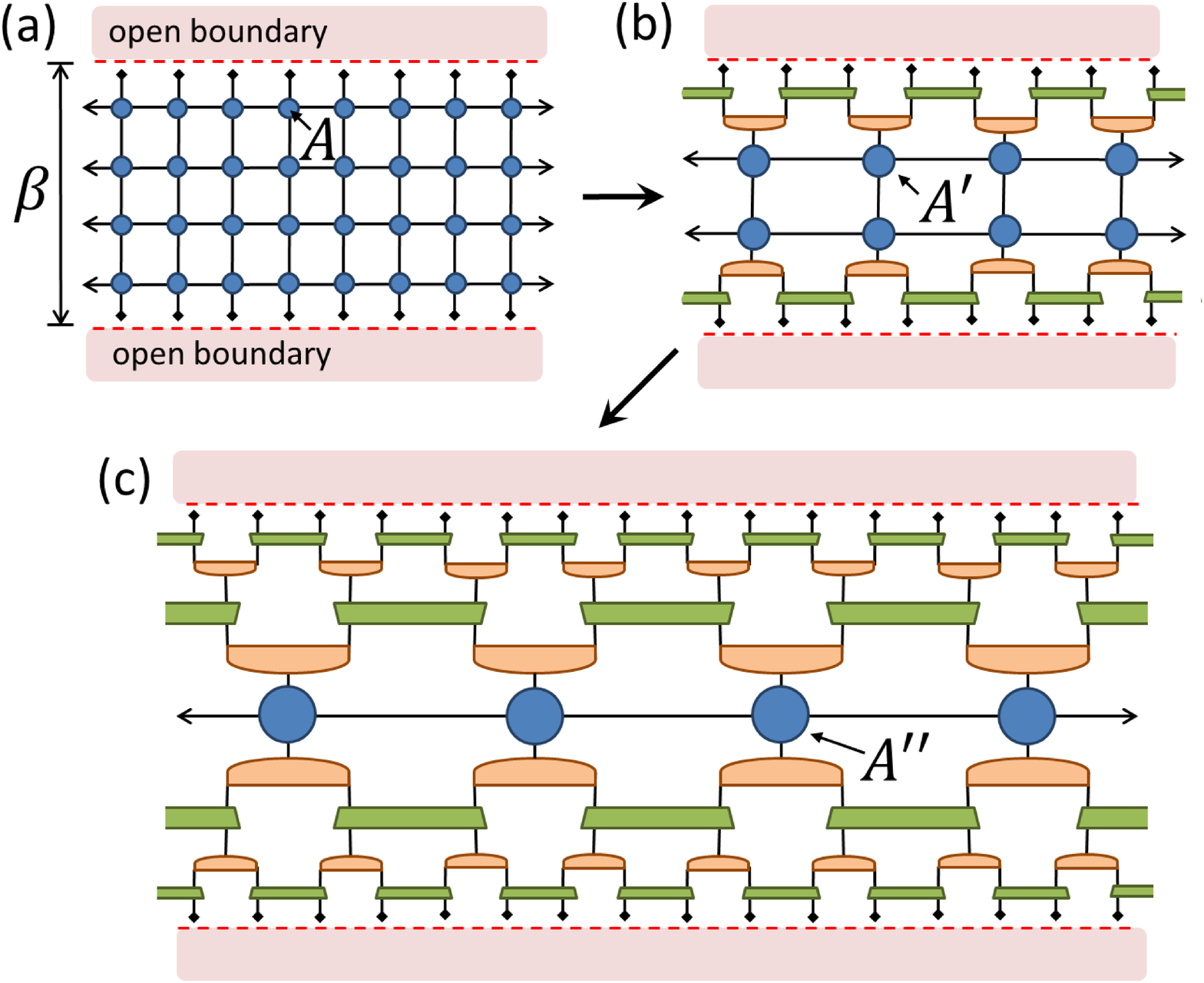}
\caption{
(a) Tensor network on an infinite strip of finite width $\beta$, with two rows of open indices. It is proportional to the thermal state $e^{-\beta H}/Z$.
(b) By coarse-graining the tensor network while leaving the open indices untouched, we obtain a new tensor network with tensors $A'$ together with an upper and lower row of disentaglers and isometries.
(c) Futher coarse-graining produces a thermal MERA.
}
\label{fig:Thermal}
\end{center}
\end{figure}

\textit{Thermal MERA.---} Let us now consider a horizontal strip of finite width $\beta$, Fig. \ref{fig:Thermal}(a), which is proportional to the thermal state $\rho_{\beta} \equiv e^{-\beta H}/Z$. This time we have two boundaries, each with an infinite row of open indices: the incoming and outgoing indices of the Euclidean time evolution operator $e^{-\beta H}$. As before, we use TNR to coarse-grain the tensor network, except near its open boundaries, where we do not touch the open indices. Fig. \ref{fig:Thermal}(b) shows the net result: a coarse-grained tensor network, with effective tensor $A'$, together with a double row of disentanglers and isometries both for the incoming and outgoing indices. After $O(\log_2(\beta))$ iterations of the coarse-graining procedure, we obtain a MERA representation of the thermal state, Fig. \ref{fig:Thermal}(c) made of $O(\log_2 (\beta))$ double layers of disentanglers and isometries for both the incoming and outgoing indices, together with a central row of tensors. Disentanglers and isometries are isometric tensors and thus do not affect the spectrum of eigenvalues of the thermal MERA, which therefore depends exclusively on the central row of tensors \cite{norm}. 

The thermal MERA obtained from TNR resembles the form first suggested by Swingle in the context of holography \cite{holography}, where a 1+1 conformal field theory is dual to a gravity theory in 3 space-time dimensions. In this context, the thermal MERA is interpreted as describing a space-like cross-section of a black hole space-time geometry. A significant difference in our construction is the central row of tensors, which is absent in Swingle's proposal \cite{holography} and provides $\rho_{\beta}$ with the correct thermal spectrum of eigenvalues $\{e^{-\beta E_i}/Z\}$, where $\{E_i\}$ are the eigenvalues of $H$. This central row of tensors can be thought of as representing the Einstein-Rosen bridge connecting the two asymptotic AdS regions \cite{Maldacena}, which seems to provide a manifestation of the ER=EPR conjecture \cite{MaldacenaSusskind}.

\begin{figure}[!t]
\begin{center}
\includegraphics[width=8cm]{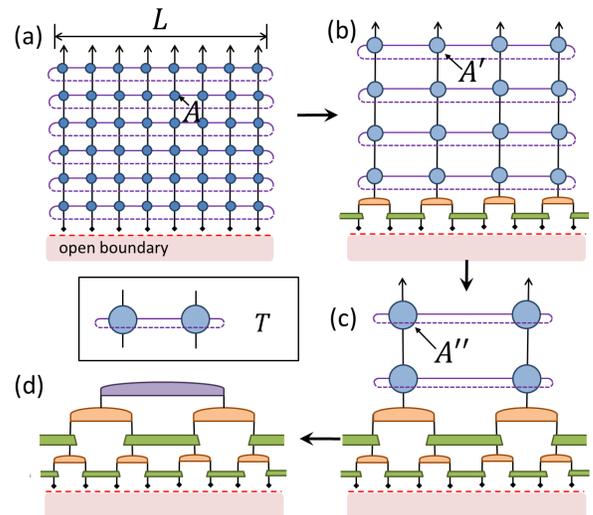}
\caption{
(a) Tensor network on a semi-infinite vertical cylinder of finite width $L$ and with a row of open indices, proportional to the ground state of $H$ on a periodic chain made of $L$ sites.
(b) Result of coarse-graining the initial tensor network while not touching its open indices. 
(c) MERA connected to a semi-infinite vertical cylinder of $O(1)$ width. 
Inset: transfer matrix $T$ of this cylinder. The eigenvectors of $T$ with largest eigenvalues correspond to the low energy eigenstates of $H$.
(d) MERA for the ground state/low energy excited states of $H$, where the top tensor is an eigenvector of the transfer matrix $T$.
}
\label{fig:FiniteSize}
\end{center}
\end{figure}

\textit{Periodic chain of size $L$.---} In our third construction, we use TNR to coarse-grain a tensor network for the ground state of Hamiltonian $H$ on a periodic chain of size $L$, see Fig. \ref{fig:FiniteSize}(a). The network consists of a semi-infinite, vertical cylinder of width $L$, with a row of open indices. After about $O(\log_2 (L))$ coarse-graining steps the size of the system has effectively become $O(1)$, see Fig. \ref{fig:FiniteSize}(c), and we have $O(\log_2 (L))$ layers of the MERA connected to a semi-infinite cylinder of $O(1)$ width. This semi-infinite cylinder can be understood as the infinite product of a transfer matrix $T$. The dominant eigenvector of $T$ leads to the ground state of $H$, whereas subdominant eigenvectors describe low energy eigenstates. 

\begin{figure}[!t]
\begin{center}
\includegraphics[width=8cm]{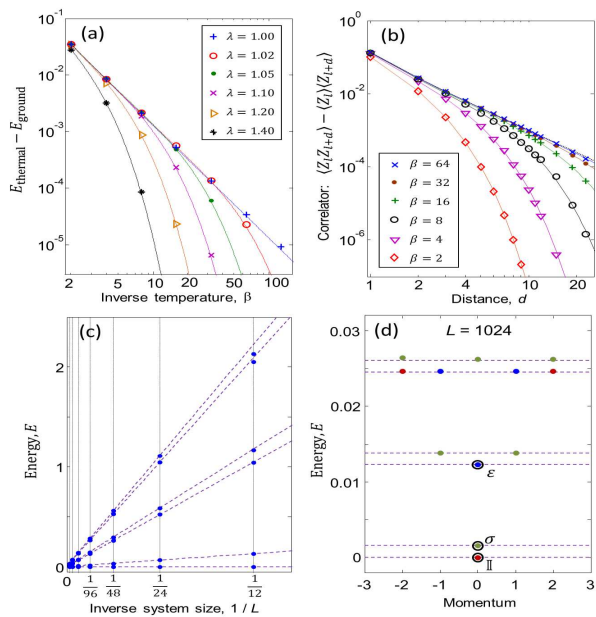}
\caption{
(a) Thermal energy per site (above the ground state energy) as a function of the inverse temperature $\beta$, for the quantum Ising model $H= \sum_i X_i X_{i+1} + \lambda \sum_i Z_i$ in an infinite chain, for different values of magnetic field $\lambda$. Continuous lines correspond to the exact solution.
(b) Connected two-point correlators at the critical magnetic field $\lambda = 1$, as a function of the distance $d$, for several values of $\beta$. Continuous lines correspond again to the exact solution.
(c) Low energy eigenvalues of $H$ for critical $\lambda=1$ as a function of $1/L$.  
(d) Low energy spectrum of $H$ for critical $\lambda = 1$ and corresponding momentum (in unites of $2\pi/L$) for $L=1024$ sites, which appear organized according to the conformal towers of the identity $\mathbb{I}$ (red), spin $\sigma$ (green), and energy density $\epsilon$ (blue) primary fields of the Ising CFT,  \cite{Henkel}. Discontinuous lines in (c) and (d) correspond to the finite-size CFT prediction, which ignores corrections of order $L^{-2}$.
}
\label{fig:ThermalPlots}
\end{center}
\end{figure}

To illustrate the computational possibilities offered by the new algorithm, we consider the one-dimensional quantum Ising model with transverse magnetic field both at finite $\beta$ for an infinite chain, and at zero temperature for a finite periodic chain of length $L$ \cite{OnlyOnce}.  The calculation required less than 5 minutes on a 2.5Ghz dual core laptop with 4Gb of memory (MERA bond dimension $\chi=10$). 
First, for $L=\infty$, Fig. \ref{fig:ThermalPlots}(a) shows the expectation value of the energy density $E_{\mbox{\scriptsize thermal}}\equiv \tr(\rho_{\beta}H)/L$ as a function of the inverse temperature $\beta$, while Fig. \ref{fig:ThermalPlots}(b) displays, at critical magnetic field, the crossover between polynomial decay of correlations at short distances (due to quantum fluctuations at criticality) and their exponential decay at longer distances (due to finite temperature statistical fluctuations). Then, for $\beta = \infty$, Figs. \ref{fig:ThermalPlots}(c)-(d) show, respectively, the low energy spectra of $H$ as a function of the inverse system size $1/L$, and the energy and momentum of low energy states for $L=1024$. In all cases, an accurate approximation to the exact result is obtained.

\textit{Discussion.---} 
Since its proposal a decade ago \cite{MERA}, the MERA has been regarded as a variational class of states. Accordingly, one attempts to approximate the ground state $\ket{\Psi}$ of $H$ by optimizing the variational parameters contained in the disentanglers and isometries of the MERA, for instance by iteratively minimizing the expectation value of $H$ \cite{algorithms1,algorithms2}. Such energy optimization is costly (it may require thousands of sweeps over scale) and prone to becoming trapped in local minima. Moreover, there is no guarantee that the end result is an approximation to the ground state $\ket{\Psi}$ -- one just has a wave-function with, hopefully, reasonably low energy. Here we have argued that, instead, an approximate MERA representation of the ground state $\ket{\Psi}$ can be obtained with TNR by coarse-graining an initial, quasi-exact tensor network representation of $\ket{\Psi}$ \cite{Suzuki}. This only requires one sweep over scale, and it is therefore computationally much more efficient \cite{AppendixD}. In addition, at each coarse-graining step TNR introduces a truncation error \cite{TNR} that can be explicitly computed. If this truncation error is sufficiently small, then one can certify that the resulting MERA approximates the true ground state $\ket{\Psi}$ within that small error \cite{AlsoBrMERA}. 

Importantly, TNR acts \textit{locally}: the coarse-graining of the tensor network at point $(x,\tau)$ only depends on the tensors in an immediate neighborhood \cite{TNR}. Let us mention three consequences for the resulting MERA. 
(i) Since TNR is not aware of the system size $L$ or inverse temperature $\beta$, it produces the same tensors $A, A', A'', \cdots$ and disentanglers and isometries for the ground state $\ket{\Psi}$ in the thermodynamic limit ($L=\beta=\infty$) as it does for the states $\rho_{\beta}$ and $\ket{\Psi^{(L)}}$ at finite $\beta$ or $L$. Thus, a single TNR calculation produces an accurate MERA approximation for all these states.
(ii) In the absence of translation invariance, where a different tensor $A_i$ may be required for each site $i$ of the lattice, the coarse-graining of different parts of the system can be conducted in parallel, leading to a massive reduction in computational time. (iii) At a conceptual level, locality of TNR implies the validity of the \textit{theory of minimal updates} \cite{MinimalUpdates1}. This theory, which asserts that only certain parts of the MERA need to be changed in order to account for a localized change in the Hamiltonian $H$ \cite{MinimalUpdates1}, is particularly useful in the study of systems with boundaries, defects, or interfaces \cite{algorithms2, Impurity}.

In Summary, in this Letter we have shown that, when applied to the Euclidean path integral restricted to several geometries, TNR \cite{TNR} produces a MERA for the ground state and thermal states of a quantum Hamiltonian, and have explored a number of consequences of this result. We conclude by briefly mentioning two more implications. 
First, thanks to this connection, TNR inherits from the MERA its ability to define an RG flow in the space of wave-functions and Hamiltonians \cite{ER,MERA,algorithms1}. Notice that extracting the emergent physics from these RG flows should be easier (both conceptually and computationally) than extracting it from the RG flow in the more abstract space of tensors, Eq. \ref{eq:TensorFlow}. 
Second, although we have focused on quantum systems, TNR can also be applied to statistical partition functions \cite{TNR}. Therefore the present construction extends the MERA formalism (including strategies to extract universal critical properties \cite{criticality1,criticality2,criticality3,criticality4,Impurity}) to classical statistical systems.

The authors thank Rob Myers for insightful comments. G. V. acknowledges support by the John Templeton Foundation and thanks the Australian Research Council Centre of Excellence for Engineered Quantum Systems. 
The authors also acknowledge support by the Simons Foundation (Many Electron Collaboration). Research at Perimeter Institute is supported by the Government of Canada through Industry Canada and by the Province of Ontario through the Ministry of Research and Innovation.


\newpage
\appendix
\vspace{2cm}
\section{Appendix A.-- A tensor network for the Euclidean path integral}
In this section we describe how to obtain a convenient, quasi-exact tensor network representation of the partition function/Euclidean time path integral $Z \equiv \tr \left( E^{-\beta H} \right)$ (or Euclidean time evolution operator $e^{-\beta H}$), starting from the short-ranged Hamiltonian $H$ of a 1D quantum system on the lattice. This tensor network is used in the main text as the starting point to obtain MERA representations for the ground state and thermal states of $H$ in infinite and finite systems, see also Appendix B. It is also used in Appendix C.

\begin{figure}[!t]
\begin{center}
\includegraphics[width=8cm]{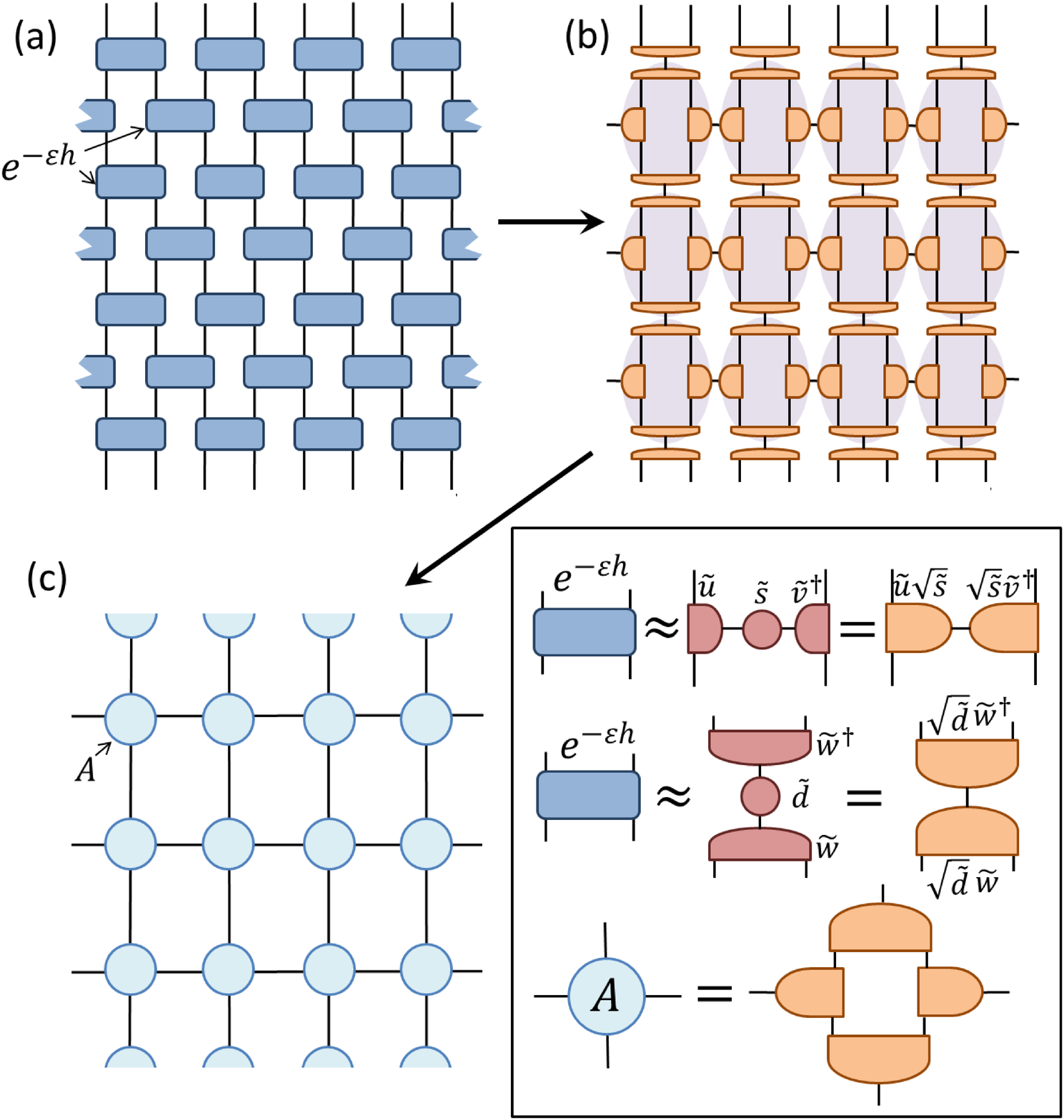}
\caption{
(a) Tensor network for the Euclidean path integral obtain by multiplying small (Euclidean time) two-site gates of the form $e^{-\epsilon h_{i,i+1}}$. 
(b) Same tensor network after decomposing each two-site gate using a singular value decomposition, according to the inset.
(c) Final tensor network for the Euclidean path integral in terms of copies of a single tensor $A$.
}
\label{fig:Suzuki}
\end{center}
\end{figure}

For simplicity, we assume that $H$ is a sum of nearest neighbor terms only,
\begin{equation}
H = \sum_i h_{i,i+1}
\end{equation}
[Longer- (but finite-) range terms can be treated with a slightly more complicated scheme.] The first step is to split $H$ into two contributions, $H = H_{\even} + H_{\odd}$ 
\begin{equation}
H_{\even} \equiv \sum_{\even ~i} h_{i,i+1},~~~~~~H_{\odd} \equiv \sum_{\odd ~i} h_{i,i+1},
\end{equation}
and use a Suzuki-Trotter decomposition \cite{Suzuki2} to approximately express $e^{\beta H}$ as the $P$-fold product of operators $e^{\epsilon H_{\teven}}$ and $e^{\epsilon H_{\todd}}$, 
\begin{equation}
e^{\beta H} \approx \left( e^{-\epsilon H_{\teven}} e^{-\epsilon H_{\todd}}\right)^P,~~~\epsilon \equiv \beta/P \ll 1.
\end{equation}
This introduces an error of order $O(\beta \epsilon)$, which therefore vanishes in the limit of small $\epsilon$/large $P$. [One can obtain an error $O(\beta(\epsilon)^n)$, $n>1$, by using a higher order Suzuki-Trotter decomposition \cite{HigherOrder}]. Since $H_{\even}$ is a sum of terms that act on different sites and therefore commute, $e^{\epsilon H_{\teven}}$ is simply a product of two-site gates, and similarly for $e^{\beta H_{\todd}}$,
\begin{equation}
e^{-\epsilon H_{\teven}} = \prod_{\even ~ i} e^{-\epsilon h_{i,i+1}},~~~ e^{-\epsilon H_{\todd}} = \prod_{\odd ~ i} e^{-\epsilon h_{i,i+1}}.
\end{equation}
Each two-site gate $e^{-\epsilon h_{i,i+1}}$ is a tensor made of four indices. Therefore we have obtained a tensor network representation of $e^{-\beta H}$, see Fig. \ref{fig:Suzuki}(a). Using singular value decompositions (in the vertical and horizontal direction, as indicated in the inset of Fig. \ref{fig:Suzuki}) we can then transform this tensor network into a new tensor network made of four-legged tensors $A$ connected according to a square lattice pattern.

\begin{figure}[!t]
\begin{center}
\includegraphics[width=8cm]{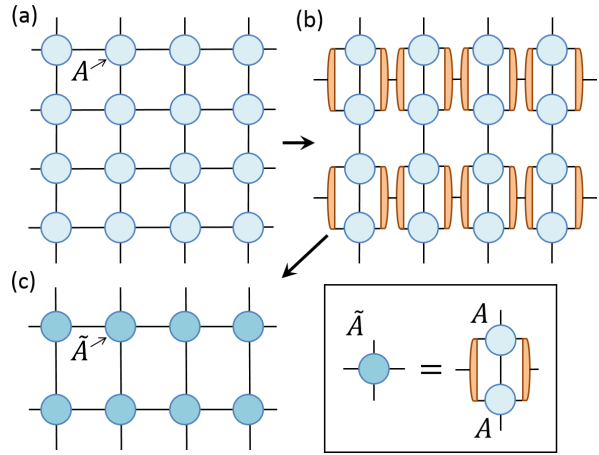}
\caption{
(a) Tensor network for the Euclidean path integral produced in Fig. \ref{fig:Suzuki}, made of copies of a single tensor $A$.
(b) We can introduce isometries (obtained e.g. from the singular value decomposition of a vertical pair of tensors $A$) to define a new tensor $\tilde{A}$.
(c) In the new tensor network, each row of tensors $\tilde{A}$ accounts for two rows of tensors $A$, in such a way that the anisotropy between space and Euclidean time directions has been reduced. The initial square tensor network discussed in the main text is the result of applying this compression in the Euclidean time directions a few times.
}
\label{fig:TimeCompression}
\end{center}
\end{figure}

However, by construction $e^{-\epsilon h_{i,i+1}}$ is very close to the identity, $e^{-\epsilon h_{i,i+1}} = I + O(\epsilon)$, which implies that the square tensor network is highly anisotropic. It is then convenient to coarse-grain in the Euclidean time direction, until space and time direction have become qualitatively equivalent, in the sense e.g. that the ordered singular values of a tensor in both directions decay roughly in the same way. This is illustrated in Fig. \ref{fig:TimeCompression}.

\section{Appendix B.--- Four geometries for an Euclidean time evolution}

In this section we review the well-known fact that the Euclidean time evolution operator $e^{-\beta H^{(L)}}$ is proportional to the thermal state $\rho_{\beta}^{(L)}$,
\begin{equation}
\rho_{\beta}^{(L)} \equiv \frac{e^{-\beta H^{(L)}}}{Z_{\beta}^{(L)}},~~~~Z_{\beta}^{(L)} \equiv  \tr \left(e^{-\beta H^{(L)}}\right),
\end{equation}
where $L$ denotes the length of a periodic chain and $\beta$ is the inverse temperature. Here we use $H^{(L)}$ and $H \equiv H^{(L\rightarrow \infty)}$ to denote a Hamiltonian on a periodic lattice of size $L$ and on an infinite lattice, respectively.

\begin{figure}[!t]
\begin{center}
\includegraphics[width=8cm]{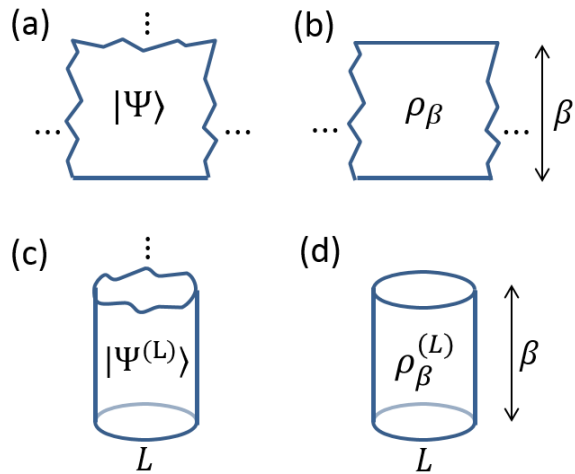}
\caption{Four geometries on which we define the tensor network for $e^{-\beta H}$. 
(a) Upper half plane $(x,\tau^{+})$, corresponding to the ground state $\ket{\Psi}$ of $H$ on an infinite lattice.
(b) Horizontal strip of finite width $\beta$, corresponding to the thermal state $e^{-\beta H}$.
(c) Semi-infinite vertical cylinder of finite width $L$, corresponding to the ground state $\ket{\Psi^{(L)}}$ of $H$ on a periodic chain of length $L$.
(d) Vertical cylinder of finite width $L$ and finite height $\beta$, corresponding to a thermal state $\rho_{\beta}^{(L)}$ on a finite periodic chain.
}
\label{fig:FourGeometries}
\end{center}
\end{figure}

We consider four geometries, obtained from combinations of infinite and finite $L$ and $\beta$. These geometries are represented in Fig. \ref{fig:FourGeometries}. The partition function $Z_{\beta}^{(L)}$ corresponds to a torus of size $L$ and $\beta$ in the horizontal and vertical directions, respectively, which are parameterized by space and Euclidean time coordinates $(x, \tau)$. Then $e^{-\beta H^{(L)}}$ corresponds to the vertical cylinder obtained by cutting the torus at $\tau=0$, as represented in Fig. \ref{fig:FourGeometries}(d). 

At infinite length $L$, this cylinder is replaced by an infinite strip of finite width $\beta$, representing $e^{-\beta H}$, which is proportional to the thermal state $\rho_{\beta}$ of Fig. \ref{fig:FourGeometries}(b), with
\begin{equation}
\rho_{\beta} \equiv \lim_{L \rightarrow \infty} \rho_{\beta}^{(L)}.
\end{equation}

At infinite inverse temperature $\beta$ (or zero temperature) we obtain a projector 
\begin{equation}
\proj{\Psi^{(L)}} = \lim_{\beta \rightarrow \infty} \frac{e^{-\beta H^{(L)}}}{Z_{\beta}^{(L)}},
\end{equation}
onto the ground state $\ket{\Psi^{(L)}}$ of $H^{(L)}$. In other words, for any state $\ket{\phi_0}$ such that $\langle \Psi^{(L)}|\phi_0\rangle \neq 0$, 
\begin{equation}
\ket{\Psi^{(L)}} \propto \lim_{\beta \rightarrow \infty} e^{\beta H^{(L)}} \ket{\phi_0}.
\end{equation}
Therefore, for finite $L$, the Euclidean time evolution operator on a semi-infinite vertical cylinder is proportional to the ground state $\ket{\Psi^{(L)}}$ of $H^{(L)}$, Fig. \ref{fig:FourGeometries}(c).

Finally, for infinite $\beta$ and $L$, the semi-infinite cylinder above is replaced with the upper half plane and produces the ground state $\ket{\Psi}$ of $H$, Fig. \ref{fig:FourGeometries}(a).

We have therefore attached a geometry to each of the ground states and thermal states $\ket{\Psi}$, $\rho_{\beta}$, $\ket{\Psi^{(L)}}$, and $\rho_{\beta}^{(L)}$. In the main text we consider explicitly the coarse-graining of the tensor network for the first three geometries in Fig. \ref{fig:FourGeometries}, that is (a) - (c), and the fourth one [the finite cylinder (d)] is left as an exercise that follows from the previous cases.

\section{Appendix C.--- Tensor Network Renormalization, step by step}

In this section we review the detailed sequence of local replacements that define the coarse-graining transformation of TNR. For convenience, we have introduced some cosmetic changes compared to Ref. \cite{TNR}.

\begin{figure}[!b]
\begin{center}
\includegraphics[width=8cm]{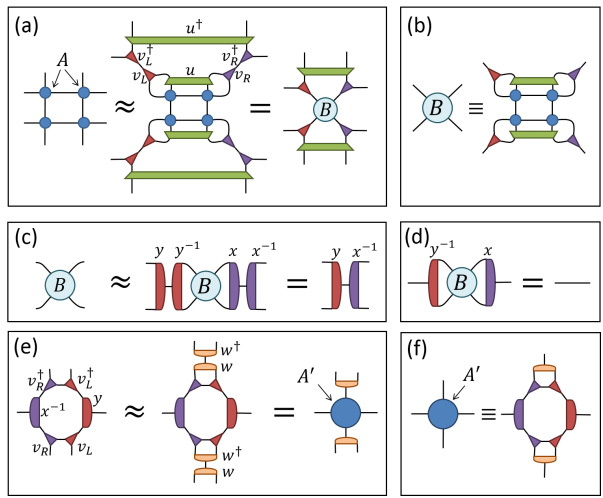}
\caption{
Sequence of elementary steps involved in TNR. 
(a) Each block of four tensors $A$ is approximately replaced with an auxiliary tensor $B$, as well as  disentanglers $u$ and isometries $v_L$ and $v_R$. The last three are determined through a local optimization \cite{TNR}, that is, one that only knows about the four tensors $A$ and is therefore insensitive to the rest of the tensor network.
(b) Definition of the auxiliary tensor $B$ in terms of $A$, $u$,$v_L$, and $v_R$.
(c) Tensor $B$ is approximately replaced by the product $y x^{-1}$.
(d) Tensors $x$ and $y$ are such that $y^{-1} B x$ is equal to the identity, and are therefore also determined locally.
(e) A combination of tensors is approximately replaced with tensor $A'$ and isometries $w,w^{\dagger}$. The isometry $w$ is determined through a local optimization.
(f) Definition of tensor $A'$. 
}
\label{fig:Steps}
\end{center}
\end{figure}

\begin{figure}[!t]
\begin{center}
\includegraphics[width=8cm]{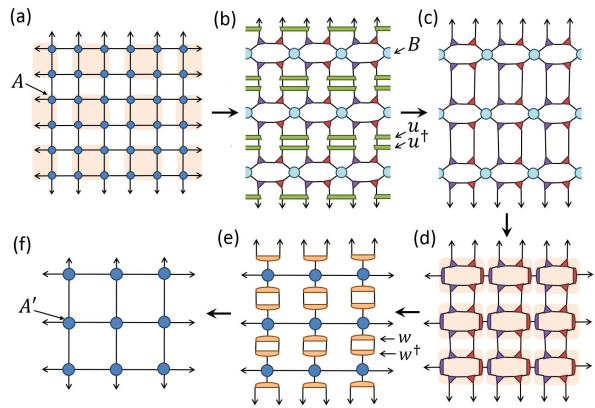}
\caption{
(a) Square tensor network made of infinitely many copies of tensor $A$. 
(b) Plaquettes of four tensors $A$ are replaced by a tensor $B$ and other tensors according to Fig. \ref{fig:Steps}(a).
(c) Pairs of disentanglers $u$ and $u^{\dagger}$ annihilate according to $uu^{\dagger}=I$.
(d) We decompose each tensor $B$ according to Fig. \ref{fig:Steps}(c).
(e) Tensor $A'$ and isometries $w$ are introduced according to Fig. \ref{fig:Steps}(e).
(f) Pairs of isometries $w$ and $w^{\dagger}$ annihilate according to $ww^{\dagger}=I$, producing the final tensor network, which contains only copies of the tensor $A'$.
}
\label{fig:Plane}
\end{center}
\end{figure}

The actual sequence is shown in Fig. \ref{fig:Steps}. It involves a number of intermediate tensors (namely $B, u, v_L, v_R, x, y, w$) in addition to the final coarse-grained tensor $A'$. When applied to an infinite square tensor network made of copies of tensor $A$, Fig. \ref{fig:Plane}, these local replacements produce a coarse-grained tensor network made of copies of a new tensor $A'$, where there is a copy of $A'$ for every four initial tensors $A$. None of the intermediate tensors appear in the coarse-grained tensor network. We can think of TNR as implementing a global change of scale, from $(x,\tau)$ to $(x',\tau')$ where
\begin{equation}
x \rightarrow x' = x/2,~~~~~~\tau \rightarrow \tau' = \tau/2.
\end{equation}

\begin{figure}[!t]
\begin{center}
\includegraphics[width=8cm]{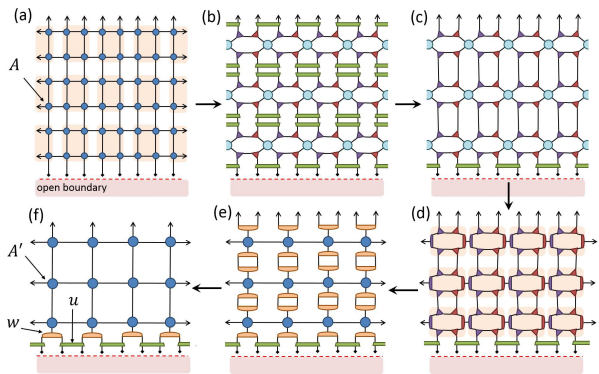}
\caption{
(a) Ground state $\ket{\Psi}$ of $H$, represented by a tensor network on the upper half plane made of copies of tensor $A$.
(b) Approximately equivalent tensor network obtained by replacing blocks of four tensors $A$ with the auxiliary tensor $B$ and disentanglers and isometries, according to Fig. \ref{fig:Steps}(a).
(c) The disentanglers annihilate pairwise according to $u u^{\dagger} = I$, everywhere except at the open boundary.
(d) We approximately replace tensor $B$ by the product $y x^{-1}$ according to Fig. \ref{fig:Steps}(c).
(e) An approximately equivalent tensor network is obtained by introducing tensor $A'$ and the isometry $w$, according to Fig. \ref{fig:Steps}(e). 
(f) The isometries annihilate pairwise according to $ww^{\dagger} = I$, everywhere except near the open boundary. The net result is thus a coarse-grained tensor network on the upper half plane with tensors $A'$ and a row of disentanglers $u$ and isometries $w$ at the open boundary, which define a layer of the MERA.
}
\label{fig:UpperHalfPlane1}
\end{center}
\end{figure}

Let us now consider instead applying TNR to coarse-grain the restriction of the initial square tensor network on the upper half plane, as represented in Fig. \ref{fig:UpperHalfPlane1}. Recall that this restriction defines a projector onto the ground state of $H$. By applying the sequence as local replacements of Fig. \ref{fig:Steps}, we see that most of the upper half plane ends up in a coarse-grained square tensor network again made of tensors $A'$. However, near the open boundary of the tensor network first a row of disentanglers $u$, and then a row of isometries $w$, accumulate. These disentanglers and isometries cannot annihilate with Hermitian conjugates $u^{\dagger}$ and $w^{\dagger}$ because those would belong to the lower half plane, which is not available. 

As discussed in the main text and Fig.1 of the main text, iteration of the coarse-graining produces a full MERA for the ground state $\ket{\Psi}$ of $H$. Let us now give a geometrical interpretation of this construction. It turns out that TNR allows to implement local scale transformations on the lattice \cite{TNRLocal}, that is, a discrete version of certain conformal transformations in the continuum.  
From this perspective, the MERA results from (the discrete version) of the conformal trasformation that maps the upper half plane into the hyperbolic plane H$^2$. 

If in the continuum we start with a flat metric in the upper half plane ($x,\tau$), the infinitesimal square line element is given by $dx^2 + d\tau^2$, which we can rewrite as $\tau^2(dx^2/\tau^2 + d\tau^2/\tau^2)$. Then a local change of scale (Weyl re-scaling) by $\tau^{-1}$ (that is, by an amount that depends on the Euclidean time coordinate $\tau$) produces a new metric
\begin{equation}
dx^2 + d\tau^2 \rightarrow (dx^2/\tau^2 + d\tau^2/\tau^2)
\end{equation}
which corresponds to H$^2$. Finally, we have also replaced the Euclidean time coordinate $\tau$ with the scale coordinate $s \equiv \log_2( \tau)$, which labels the different layers of the MERA.

Returning to the discrete geometry of the square tensor network, notice that the net effect of iterating TNR on the (discrete) upper half plane as discussed above is that we are introducing a change of scale where locally the metric is changed by an amount proportional to the distance $\tau = 2^s$ to the boundary, that is exponential in the scale variable s, as in the above conformal transformation.

\begin{figure}[!htb]
\begin{center}
\includegraphics[width=8cm]{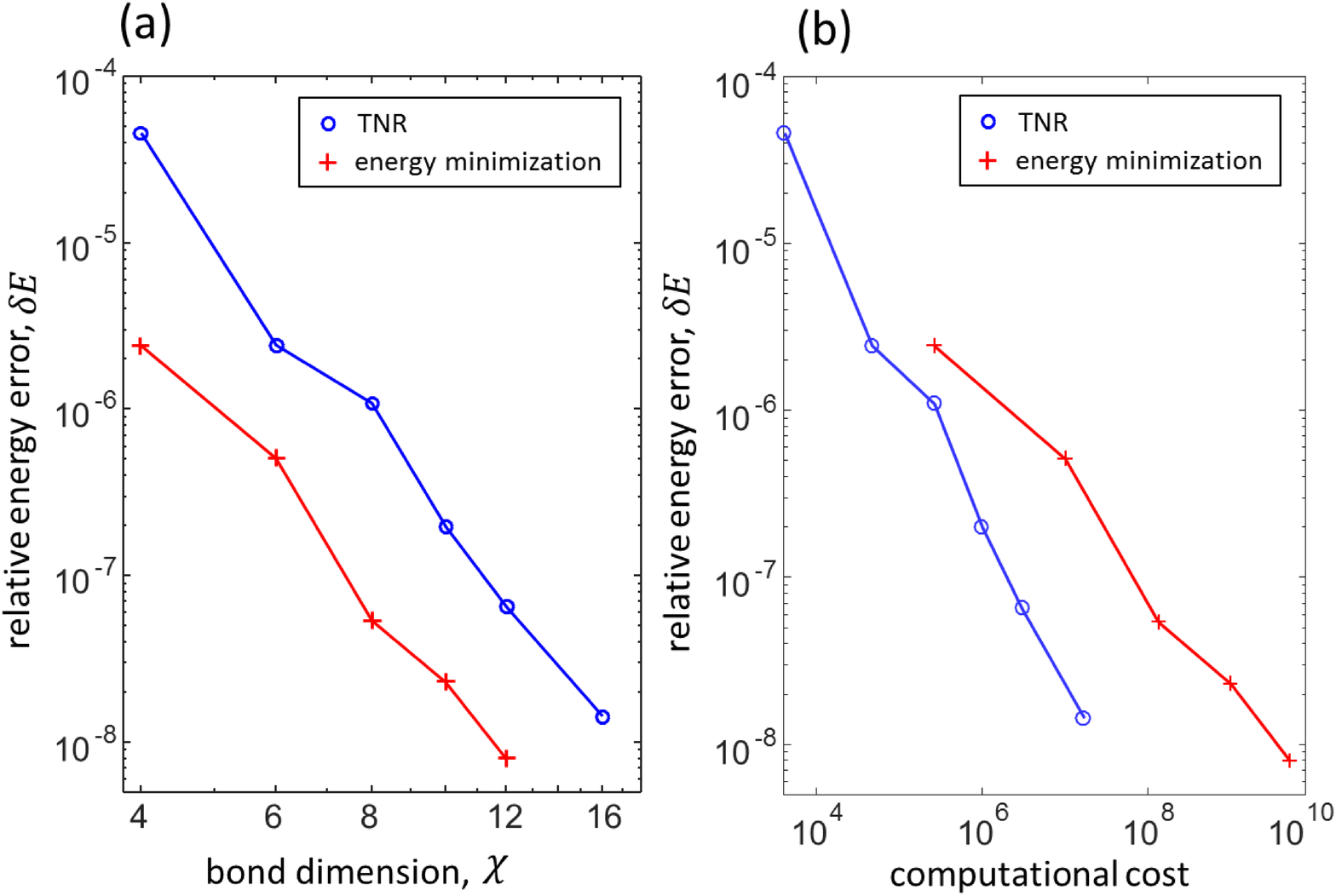}
\caption{(a) Relative error in the ground energy (per-spin) for the ground state of the $1D$ quantum Ising model at criticality, comparing MERA of bond dimension $\chi$ optimized either using TNR or variational energy minimization. (b) Relative error in the ground energy from (a), now plotted as a function of the leading order computational cost of the optimization method, which scales as $O(\chi^6)$ for TNR and $O(\chi^9)$ for energy minimization.}
\label{fig:EnergyCompare}
\end{center}
\end{figure}
 
\section{Appendix D.-- TNR versus energy minimization}

In this section we compare the accuracy and efficiency of two methods that allow us to obtain a MERA approximation to the ground state $\ket{\Psi}$ of a many-body Hamiltonian $H$: (i) TNR and (ii) energy minimization \cite{algorithms1,criticality2,criticality3}. We use the $1D$ quantum Ising model, with Hamiltonian
\begin{equation}
H= \sum_i X_i X_{i+1} + \lambda \sum_i Z_i,
\end{equation}
where $X$ and $Z$ are Pauli matrices, as a benchmark model for comparison. The ground state of the Ising Hamiltonian $H$ at critical magnetic field strength, $\lambda = 1$, is represented using a scale-invariant MERA obtained in these two ways. The scale invariant MERA considered here consists of three transitional layers followed by infinitely many copies of a scale-invariant layer \cite{algorithms1,criticality2,criticality3}.

\textit{TNR.---} In the first approach, we start by generating the Euclidean path integral for $H$, as described in section A. Then we apply four coarse-graining iterations of TNR, after which the flow in the space of tensors has reached an approximate scale-invariant fixed-point $A_\textrm{fp}$, see Eq. 1 in the main text. A scale-invariant MERA is then built from tensors (disentanglers and isometries) produced while implementing TNR, as described in the main text. Here the first three iterations of TNR generate the three transitional layers, while the fourth iteration of TNR is used to build the scale-invariant layer. 

\textit{Energy minimization.---} In the second approach, we use the energy minimization algorithm of Ref. \cite{algorithms1,criticality2,criticality3} to directly optimize a scale-invariant MERA (with three transitional layers preceding the scale-invariant layers) by iteratively minimizing the expectation value of the $H$.

The comparison between the ground energy error $\delta E$ obtained by the two methods is presented in Fig. \ref{fig:EnergyCompare}. For an equivalent bond dimension $\chi$, energy minimization produces a MERA with smaller error $\delta E$ than TNR, by a factor $k \approx 10$ that is roughly independent of $\chi$. That energy minimization would produce a more accurate estimate of the ground state energy than TNR was to be expected, for two reasons. First, the energy minimization algorithm directly targets the ground energy as its figure of merits, whereas TNR does not. Second, the energy minimization algorithm optimizes the MERA by sweeping back and forth over length scales/layers many times, whereas TNR considers each length scale only once, without returning back to optimize shorter length scales after coarse-graining larger length scales. 

However the cost of optimization also differs between the approaches: the leading order cost of energy minimization scales as $O(\chi^9)$, whereas the cost of TNR scales as $O(\chi^6)$. There is also a significant difference in the number of optimization steps required to reach a converged result. The energy minimization algorithm requires many more steps to converge, as the algorithm sweeps over scales thousands of times, compared to TNR, which treats each scale only once. The difference between the two approaches in terms of the number of optimization steps required is difficult to quantify as it depends on many factors, but here we find that energy minimization requires roughly an order of magnitude more steps than TNR. Fig. \ref{fig:EnergyCompare}(b) compares the energy error $\delta E$ between energy minimization and TNR as a function of the leading order computational costs of the approaches. Even though this comparison does not take into account the difference in the number of optimization steps between the approaches, TNR is already seen to be more computationally efficient in producing a MERA approximation to the ground state within some specified accuracy $\delta E$ than the energy minimization algorithm.

\section{Appendix E.-- TRG yields a tree tensor network (TTN)}

The main result of this paper is to show that when TNR is applied to a tensor network representing the Euclidean path integral of a quantum Hamiltonian $H$ (restricted to the upper-half plane), it generates a MERA approximation for the ground state of $H$. In this section we show an analogous result for TRG: we show that when applied to the same tensor network, it generates instead a tree tensor network (TTN) approximation for the ground state of $H$.

This result, which extends to several generalizations of TRG, offers an alternative perspective to understand the qualitative differences between TRG and TNR, and provides a robust theoretical argument for why TRG cannot realize a critical fixed-point tensor while TNR can \cite{TNR}. Indeed, the failure of TRG to produce a critical fixed-point tensor can be understood to follow from the well-known fact that the TTN ansatz (with fixed bond dimension) cannot reproduce the logarithmic scaling of entanglement entropy, $S_L \sim \log(L)$, found in the ground states of $1D$ quantum systems at a critical point. 

\begin{figure}[!htb]
\begin{center}
\includegraphics[width=8cm]{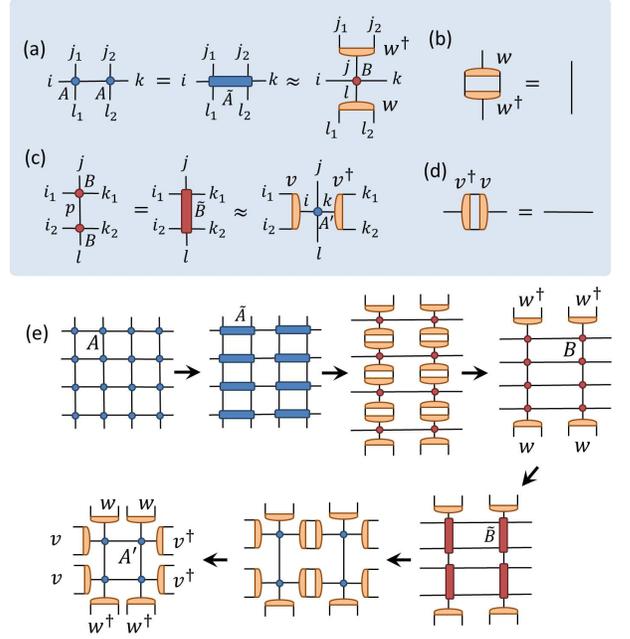}
\caption{
(a) A left-right pair of tensors $A$ is multiplied to form a new tensor $\tilde A$, which is then decomposed using a truncated HOSVD into the product of a core tensor $B$ and a pair of isometries $w$, see also Eqs. \ref{eq:E1} and \ref{eq:E2}. 
(b) The product of an isometry $w$ and its conjugate produces the identity, see Eq. \ref{eq:E3}. 
(c) A up-down pair of tensors $B$ is multiplied to form a new tensor $\tilde B$, which is then decomposed using a truncated HOSVD into the product of a core tensor $A'$ and a pair of isometries $v$. 
(d) The product of an isometry $v$ and its conjugate produces the identity. 
(e) The sequence of transformations from (a-d) above is used to coarse grain a $4\times 4$ network of tensors $A$, yielding a $2\times2$ network of tensors $A'$ surrounded by isometries $w$ and $v$.}
\label{fig:HOSVD}
\end{center}
\end{figure}

For simplicity, we will derive this result not for the original TRG, but for an improved version known as \textit{higher order tensor renormalization group} (HOTRG), which was proposed in Ref.\cite{HOTRG}. However, the same conclusions can also be shown to hold for the square-lattice form of the original TRG algorithm proposed in Ref.\cite{TRG}. We begin with brief summary of HOTRG, see Fig. \ref{fig:HOSVD}.

\begin{figure}[!htb]
\begin{center}
\includegraphics[width=8cm]{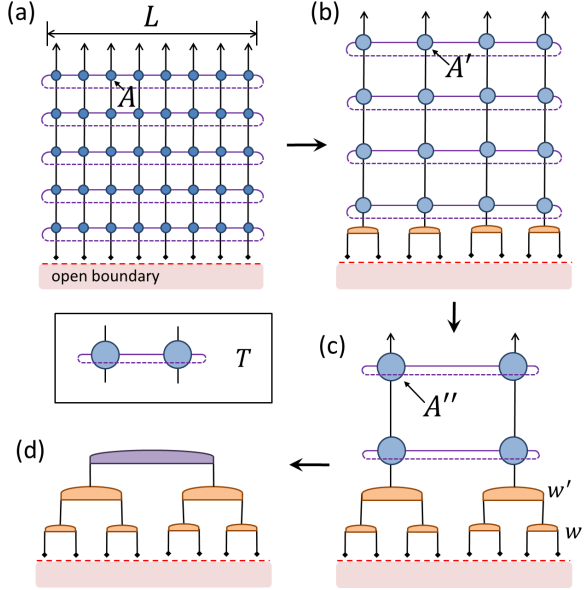}
\caption{
(a) Square lattice tensor network representing the ground state of a quantum Hamiltonian $H$ on a ring of $L$ sites through an infinite Euclidean time evolution. The tensor network describes a semi-infinite vertical cylinder of width $L$ and with a row of open indices. 
(b) The result of coarse-graining the initial tensor network with one iteration of  HOTRG, see Fig. \ref{fig:HOSVD}(e), while not touching the $L$ open indices at the bottom. 
(c) After coarse-graining with a second iteration of HOTRG, a TTN connected to a semi-infinite vertical cylinder of $O(1)$ width is produced. Inset: transfer matrix $T$ of this cylinder. The eigenvectors of $T$ with largest eigenvalues correspond to the low energy eigenstates of $H$. 
(d) TTN for the ground state/low energy excited states of $H$, where the top tensor is an eigenvector of the transfer matrix $T$.}
\label{fig:TTN}
\end{center}
\end{figure}

Consider a square lattice tensor network made of copies of a tensor $A$ with components $A_{ijkl}$ and bond dimension $\chi$. In HOTRG, one coarse-grains the tensor network first in one direction and then in the other. We shall begin with the horizontal axis. In the first step, left-right pairs of tensors $A$ are multiplied together, obtaining the six-index tensor $\tilde A$,
\begin{equation}
{\tilde A_{i,{j_1},{j_2},k,{l_1},{l_2}}} \equiv \sum\limits_p {{A_{i,{j_1},p,{l_1}}}{A_{p,{j_2},k,{l_2}}}}, \label{eq:E1}
\end{equation}
as shown in Fig.\ref{fig:HOSVD}(a). We now should truncate the double indices $j\equiv (j_1 j_2)$ and $l\equiv (l_1 l_2)$ of $\tilde A$, which are of dimension $\chi^2$, down to dimension $\chi$. In HOTRG this truncation is performed using a generalized form of the SVD called the higher order singular value decomposition (HOSVD), a method of decomposing an $N$ rank tensor into the product of a core tensor and a set of isometric tensors. Using the HOSVD, the tensor $\tilde A$ is approximately decomposed into a core tensor $B$ (with all four indices of dimension $\chi$), and a pair of isometries $w$ and $w^\dag$ (with their three indices also of dimension $\chi$), 
\begin{equation}
{{\tilde A}_{i,{j_1},{j_2},k,{l_1},{l_2}}} \approx \sum\limits_{j,l = 1}^\chi  {{w_{l,{l_1},{l_2}}}{B_{i,j,k,l}}{{({w_{j,{j_1},{j_2}}})}^ * }}, \label{eq:E2}
\end{equation}
see also Fig. \ref{fig:HOSVD}(a). [Note that we are assuming, for simplicity, that the network is invariant under spatial reflection along the horizontal axis such that the top and bottom isometries $w$ are conjugates of one another, although the same result could also be reached without this simplifying assumption.] Isometries $w$ satisfy the following relation, 
\begin{equation}
\sum\limits_{{j_1},{j_2} = 1}^\chi  {{w_{l,{j_1},{j_2}}}{{({w_{j,{j_1},{j_2}}})}^ * }}  = {\delta _{l,j}},
\end{equation}
as depicted in Fig.\ref{fig:HOSVD}(b). After the HOSVD, we obtain a new square lattice tensor network composed of tensors $B$ and rescaled by a factor of 2 in the horizontal direction. Following the same steps but now in the vertical direction, up-down pairs of tensors $B$ are multiplied together to form a six-index tensor $\tilde B$, which can then be decomposed using the HOSVD into a product of a core tensor $A'$ and a pair of isometries $v$ and $v^\dag$, see Fig.\ref{fig:HOSVD}(c). [Again, for simplicity, we have assumed that the network is invariant under appropriate spatial reflection.] The combination of horizontal and vertical contraction steps has taken the initial square lattice network of tensors $A$ to a new network of tensors $A'$, one which is rescaled by a factor of 2 on both axes. This constitutes a single iteration of the HOTRG approach. By iteration, we generate a sequence of increasingly coarse-grained lattices defined in terms of tensors,
\begin{equation}
A\mathop  \to \limits^{\left( {w,v} \right)} A'\mathop  \to \limits^{\left( {w',v'} \right)} A''\mathop  \to \limits^{\left( {w'',v''} \right)} A''' \to \cdots .\label{eq:E3}
\end{equation}

Let us now apply HOTRG to a square lattice tensor network representation of the ground state of a Hamiltonian $H$ on a periodic chain of size $L$, as we did in the main text for TNR. The network consists of a semi-infinite, vertical cylinder of width $L$, with a row of open indices at the bottom, see Fig. \ref{fig:TTN}(a). Each coarse-graining iteration rescales the bulk of the network by a linear factor of 2, while leaving a row of isometries $w$ along the open boundary, see Fig.\ref{fig:TTN}. After about $\log_2 (L)$ coarse-graining steps the size of the system has effectively become $O(1)$, and we have $O(\log_2 (L))$ layers of the TTN connected to a semi-infinite cylinder of $O(1)$ width. This semi-infinite cylinder can be understood as the infinite product of a transfer matrix $T$; taking the dominant eigenvector of $T$ as the top tensor of the TTN leads to the ground state of $H$. Thus the HOTRG approach, applied to the Euclidean path integral of a quantum system $H$, has been shown to generate a TTN approximation for its ground state.

Since the introduction of TRG, a significant improvement in renormalization schemes for tensor networks has been the use of the so-called environment to achieve a more accurate truncation. The second renormalization group (SRG) proposed in Ref.\cite{SRG} uses to knowledge from the environment, obtained approximately by iteratively sweeping over all length scales, to allow the tensor truncations to be performed so as to minimize the \textit{global} error, as opposed to the \textit{local} error minimized in TRG and HOTRG. Ref. \cite{HOTRG} introduced a version of SRG based upon HOSVD, called the HOSRG, which we will now consider. The use of the environment allows HOSRG to more accurately discern which degrees of freedom can be truncated, and which should be retained, at each RG step. This allows for a smarter choice of isometries. However, HOSRG is structurally still based upon the same blocking scheme as HOTRG, Fig.\ref{fig:HOSVD}(e). It follows that HOSRG also leads to a TTN representation of ground and excited states as depicted in Fig.\ref{fig:TTN}. The only difference being that the isometries $w$ that compose the TTN have been better chosen with HOSRG than with HOTRG, as they minimize the global error. In particular, for the same bond dimension $\chi$, HOSRG may produce a more accurate TTN approximation to the ground state than HOTRG.

However, as it happens with HOTRG (and TRG), HOSRG cannot generate a proper RG flow in the space of tensors -- that is, one that at criticality flows to a critical fixed-point tensor. Indeed, HOSRG produces a TTN, which is known to only be able to accurately describe a critical system if the bond dimension grows with increasing scale. But a bond dimension that grows with scale is incompatible with having a fixed-point tensor.


\begin{thebibliography}{99}
\bibitem{nonTI} We assume translation invariance. For a system that is not invariant under translations, a different tensor $A_i$ may be needed for each site $i$ of the lattice.
\bibitem{HoloThermal} Previous proposals for a thermal MERA discussed in Refs. \cite{holography} in the context of holography have a flat spectrum of eigenvalues and therefore do not reproduce the correct, finite temperature spectrum of $\rho_{\beta}$.
\bibitem{layer} On the plane, the double row of disentanglers and isometries will annihilate with a double row of conjugate disentanglers and isometries corresponding to $\tau < 0$. Instead, on the half upper plane, these tensors remain.
\bibitem{norm} The normalization factor $Z$ corresponds to the trace of the thermal MERA. Since the conjugate layers of disentanglers and isometries do not change this trace, $Z$ is also the trace of the central row of tensors $A^{\max}$, or an infinite power of some (transfer) matrix $M$ obtained from tensor $A^{\max}$ by contracting its vertical indices. A normalized thermal MERA is then obtained by replacing $A^{\max}$ with $A^{\max}/\lambda$, where $\lambda$ is the dominant eigenvalue of $M$.
\bibitem{Suzuki} The initial tensor network representation of $\ket{\Psi}$ is quasi-exact, in that the Suzuki-Trotter decomposition used to produce it (Appendices A and B) introduces only small errors, which can be systematically reduced.
\bibitem{AlsoBrMERA} We emphasize, however, that TNR is not committed to exclusively producing a MERA. For instance, if at some scale $s$ the ground state approximately decomposes as the product of two ground states, then the corresponding tensor $A^{(s)}$ will approximately factorize as $A_1^{(s)}\otimes A_2^{(s)}$, and TNR will give rise instead to a branching MERA \cite{branchingMERA}, which was proposed to describe e.g. highly entangled phases of matter in $D\geq 2$ dimensions.
\end{thebibliography}
\end{document}